\documentclass[%
    aip,
    cha,
    reprint,
    amsmath,
    amssymb,
    floatfix,
    showkeys,
]{revtex4-2}

\draft
\usepackage{hyperref}
\usepackage{mathtools}
\usepackage{xcolor}

\newcommand{\w}{\omega}
\newcommand{\e}{\varepsilon}
\renewcommand{\d}{\mathrm{d}}

\renewcommand{\P}{\mathbb P}
\newcommand{\Cont}{\text{cont}}
\newcommand{\Exp}{\text{exp}}

\begin{document}

\preprint{CHA/123-QED}

\title{Noise-induced chaos: a conditioned random dynamics perspective}

\author{Bernat Bassols-Cornudella}
\email[Corresp. author: ]{bernat.bassols-cornudella20@imperial.ac.uk}
\affiliation{Department of Mathematics, Imperial College London, London SW7 2AZ, UK}
\author{Jeroen S.W. Lamb}
\email[Email address: ]{jeroen.lamb@imperial.ac.uk}
\affiliation{Department of Mathematics, Imperial College London, London SW7 2AZ, UK}
\affiliation{International Research Center for Neurointelligence, The University of Tokyo, Tokyo 113-0033, Japan}
\affiliation{Centre for Applied Mathematics and Bioinformatics, Department of Mathematics and Natural Sciences, Gulf University for Science and Technology, Halwally 32093, Kuwait}

\date{\today}

\begin{abstract}
We consider transitions to chaos in random dynamical systems induced by an increase of noise amplitude. We show how the emergence of chaos (indicated by a positive Lyapunov exponent) in a logistic map with bounded additive noise can be analyzed in the framework of conditioned random dynamics through expected escape times and conditioned Lyapunov exponents for a compartmental model representing the competition between contracting and expanding behavior. In contrast to the existing literature, our approach does not rely on small noise assumptions, nor refers to deterministic paradigms.
We find that the noise-induced transition to chaos is caused by a rapid decay of the expected escape time from the contracting compartment, while all other order parameters remain approximately constant.
\end{abstract}


\maketitle

\begin{quotation}
    It is well-known that adding noise to a nonlinear system with a non-chaotic attractor can cause the attractor to become chaotic. This phenomenon is known as noise-induced chaos. In the literature, studies of such bifurcations have focused on asymptotic properties of order parameters near criticality under small noise assumptions, with reference to deterministic paradigms. We consider noise-induced chaos from a novel perspective, employing concepts from the theory of conditioned random dynamical systems, such as the recently proposed conditioned Lyapunov exponent, that exist independently of the strength of the noise. In particular, we find that in a prototypical random logistic map, the noise-induced transition to chaos is caused by a noise-induced (rapid) decrease of the average escape time from the original attractor. This paper successfully pilots conditioned dynamics as a mathematical framework to study bifurcations in random systems beyond the small noise setting.
\end{quotation}

\section{Introduction}

Chaos theory is considered to be one of the scientific revolutions of the 20th century, providing new paradigms for the understanding of predictability and stability with broad repercussions, from  fundamental science to applied engineering. 

A key question in this theory addresses the mechanisms through which chaos arises. In low-dimensional deterministic systems, various universal routes to chaos have been identified, e.g. via period-doubling cascades\cite{Feigenbaum1978}, intermittency\cite{Pomeau1980}, torus bifurcations\cite{Ruelle1971} and homoclinic bifurcations\cite{Shilnikov2001}.

There is recently a growing recognition of the importance to study dynamics in the presence of noise, for instance in the context of quantum mechanics\cite{Schmolke2022}, turbulence\cite{Bedrossian2022} and ecology\cite{Meng2020}. Such random dynamical systems display behaviors that are different from their deterministic counterparts.\cite{Horsthemke1984, Arnold1998} The corresponding theory, however, remains in its early stages of development. In particular, noise-induced transitions such as noise-induced order\cite{Matsumoto1983, Galatolo2020} (where adding noise to a deterministic chaotic systems suppresses chaos) and noise-induced chaos\cite{Crutchfield1980} (where adding noise to a non-chaotic deterministic system causes chaos) have been long recognized but remain poorly understood.\cite{Lai2011}

In this paper, we study noise-induced chaos using concepts from conditioned random dynamical systems, providing the first practical application of the recently introduced conditioned Lyapunov exponent.\cite{Engel2019, Castro2022} 

Previous studies on noise-induced chaos\cite{Crutchfield1980, Grebogi1983, Kantz1985, Grebogi1988, Lai2003, Tel2008, Lai2011} have primarily focused on asymptotic properties of order parameters near criticality, relying on certain insights into long chaotic transients in deterministic systems and relating escape times to fractal properties of the repeller.\cite{Tel2008} Consequently, these studies assume a perturbative small noise setting. Nonetheless, noise-induced chaos arises only beyond a determined finite noise strength, with markedly different dynamics at smaller noise levels. The interplay between noise and underlying deterministic equations of motion lies at the heart of random dynamical systems theory. While in the limit of small noise the resulting dynamics is almost deterministic, it is beyond this limit where truly novel dynamical phenomena, like noise-induced chaos, emerge.

This pilot study provides motivation for further research into the development of effective compartmental models for random dynamical systems. Despite progress in the theory of conditioned random dynamics\cite{Engel2019,Castro2021,Castro2022,Castro2023b}, further theoretical progress is necessary to gain deeper insights. Similarly, efficient computational tools for the  analysis of concrete examples remain only in their early stages of development.

\section{Random logistic map}

As a paradigmatic prototypical model for noise-induced chaos, we consider a logistic map with uniform bounded additive noise:
\begin{equation}\label{eq:rlm}
\begin{split}
    x_{n+1} &= f(x_n) + \w_n \eqqcolon f_{\w_n}(x_n),\\
    \w_n &\sim \P \coloneqq \text{Unif}(-\e, \e),
\end{split}
\end{equation}
where $f(x) \coloneqq 3.83x(1-x)$ is the logistic map with growth factor $3.83$ and $\e>0$ is the noise amplitude. In the absence of noise ($\e=0$) almost all initial conditions in $[0,1]$ are attracted to a period-3 orbit while the complement of the domain of attraction contains a chaotic repeller. At each iteration, $\w_n$ is chosen i.i.d. from $[-\e, \e]$ with uniform probability $\P$. We refer to the semi-infinite sequence $\w = \w_0\w_1\dots$ as a noise realization and write $f^{n}_\w(x_0) \coloneqq f_{\w_{n-1}} \circ \ldots \circ f_{\w_0}(x_0).$

Eq.~\eqref{eq:rlm} was considered in Ref.~\onlinecite{Crutchfield1982} with unbounded Gaussian  noise. In many modeling settings, bounded noise is natural\cite{dOnofrio2013} and unbounded noise may have unintended dynamical consequences. For instance, with Gaussian unbounded noise, trajectories of Eq.~\eqref{eq:rlm} almost surely tend to $-\infty$, whereas the bounded noise version has a bounded attractor for modest noise amplitudes. As a result, numerical results in Ref.~\onlinecite{Crutchfield1982} concern empirical statistics of finite time observations, which are transient.

Eq.~\eqref{eq:rlm} is known to exhibit three dynamical regimes depending on the value of the noise amplitude\cite{Sato2018}, $\e$, which are characterized by the support of the (unique) stationary density, $p$, and the sign of the Lyapunov exponent, $\Lambda$. Recall that the Lyapunov exponent is an indicator of a system's stability in the sense that it measures the exponential rate of expansion (when positive) or contraction (when negative) of the distance between forward orbits of two infinitesimally close points\cite{Viana2014}. For the random logistic map with bounded additive noise in Eq.~\eqref{eq:rlm}, the Lyapunov exponent is given by\cite{Sato2018}:
\[%\label{eq:lyap_exp}
\begin{split}
    \Lambda &\coloneqq \lim_{N \to \infty} \frac{1}{N} \sum_{n = 0}^{N-1} \ln|a(1 - 2x_n)|\\
    &= \int_A \ln|a(1 - 2y)|p(y)\d y,
\end{split}
\]
where the latter equality holds for $p$-almost every $x_0$ by Birkhoff's ergodic theorem. Here, $A = \mathrm{supp}\ p$ is the unique attractor of the random logistic map identified in Ref.~\onlinecite{Mayer-Kress1981}.

\begin{figure*}[hbt]
    \centering
    \includegraphics[width = .8\textwidth]{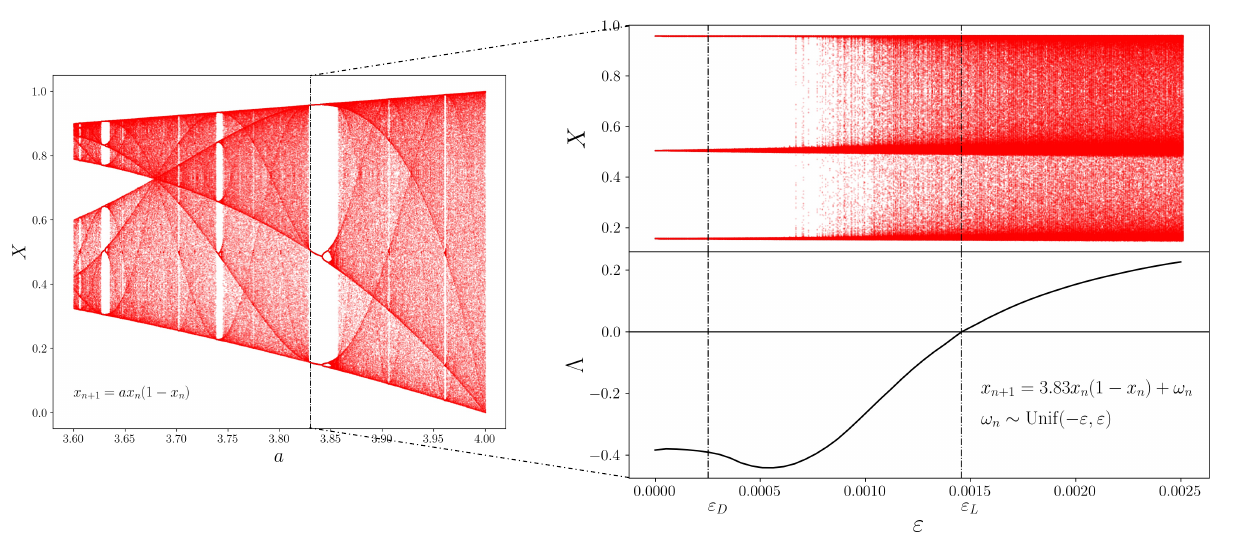}
    \caption{On the left, bifurcation diagram for the deterministic logistic map $x_{n+1} = ax_n(1-x_n)$, with $a \in [3.6,4]$. The value $a = 3.83$ shows to be inside the three periodic window, where almost all orbits converge towards a three periodic orbit. On the right, bifurcation diagram for the random logistic map $x_{n+1} = 3.83x_n(1-x_n) + \w_n$, together with the system's Lyapunov exponent, $\Lambda$, with $\w_n \sim \text{Unif}(-\e, \e)$ and $\e \in [0,0.0025]$. The three phases are separated by two dashed vertical lines at the bifurcation points $\e_D$ and $\e_L$.}
    \label{fig:bif_diagram}
\end{figure*}

The limiting dynamics of this system are sketched in FIG.~\ref{fig:bif_diagram}. We briefly summarize the main features of interest from Refs.~\onlinecite{Mayer-Kress1981, Sato2018}. In the absence of noise, $\e = 0$, almost all orbits converge towards a period three attractor. If noise is small, $\e< \e_D \approx 0.000251$, this attracting cycle fattens into a random periodic attractor $A$, consisting of three disjoint intervals that are cyclically permuted by the dynamics. Orbits in this attractor uniformly converge to each other if and only if they start in the same interval. 

As the noise level $\e$ increases, at $\e_D$, a topological bifurcation merges these three intervals into one connected attractor, after which any two orbits in this attractor almost surely converge to a single random point attractor. This topological bifurcation is naturally viewed from a set-valued perspective\cite{Lamb2015} and associated with a saddle-node bifurcation of a map with extreme noise realization, cf.~Ref.~\onlinecite{Sato2018} in the context of Eq.~\eqref{eq:rlm}. In this regime, noise-induced synchronization (in the sense of almost sure convergence of trajectories) is ensured\cite{Newman2018} by the negativity of the Lyapunov exponent $\Lambda$ although the convergence to the random fixed point is no longer uniform.\cite{Callaway2017} At $\e=\e_L \approx 0.00146$, the Lyapunov exponent becomes positive and the random point attractor bifurcates into a random chaotic attractor. 

Following Ref.~\onlinecite{Sato2018}, we refer to the three corresponding dynamical regimes as phase I ($\e < \e_D$), phase II ($\e_D < \e < \e_L$) and phase III ($\e_L < \e$).

In this paper we focus on the transition from (non-uniform) synchronization in phase II ($\e_D < \e < \e_L$) to chaos in phase III $(\e_L<\e)$. To reveal the mechanism driving this transition, we consider the two-point motion of pairs of initial conditions $(x_0, y_0)$ under the random dynamics with identical noise realization:
\[%\label{eq:tpm}
    \begin{array}{ccccc}
        x_{n+1} &=& ax_n(1-x_n) &+& \w_n,\\
        y_{n+1} &=& ay_n(1-y_n) &+& \w_n.
    \end{array}
\]
To track the relative exponential expansion and contraction we introduce the coordinates 
\[%\label{eq:log_coords}
    u_n \coloneqq \ln |x_n - y_n| 
    \quad \text{and} \quad v_n \coloneqq \frac{x_n + y_n}{2},
\]
in terms of which we rewrite the two-point motion as
\[%\label{eq:tpm_log}
    \begin{array}{ccl}
    u_{n+1} &=& u_n + \ln|a(1-2v_n)|,\\
    v_{n+1} &=& av_n(1-v_n) + \w_n - \dfrac{a}{4} \exp(2u_n).
    \end{array}
\]
The benefit of this change of coordinates is two-fold. First, it allows for accurate computation of fluctuations of the distance between two trajectories close to synchronization. Secondly, exponential growth or decay of the distance between two orbits manifests itself as linear growth or decay of $u_n$.

\begin{figure}[hbt]
    \centering
    \includegraphics[width = .48\textwidth]{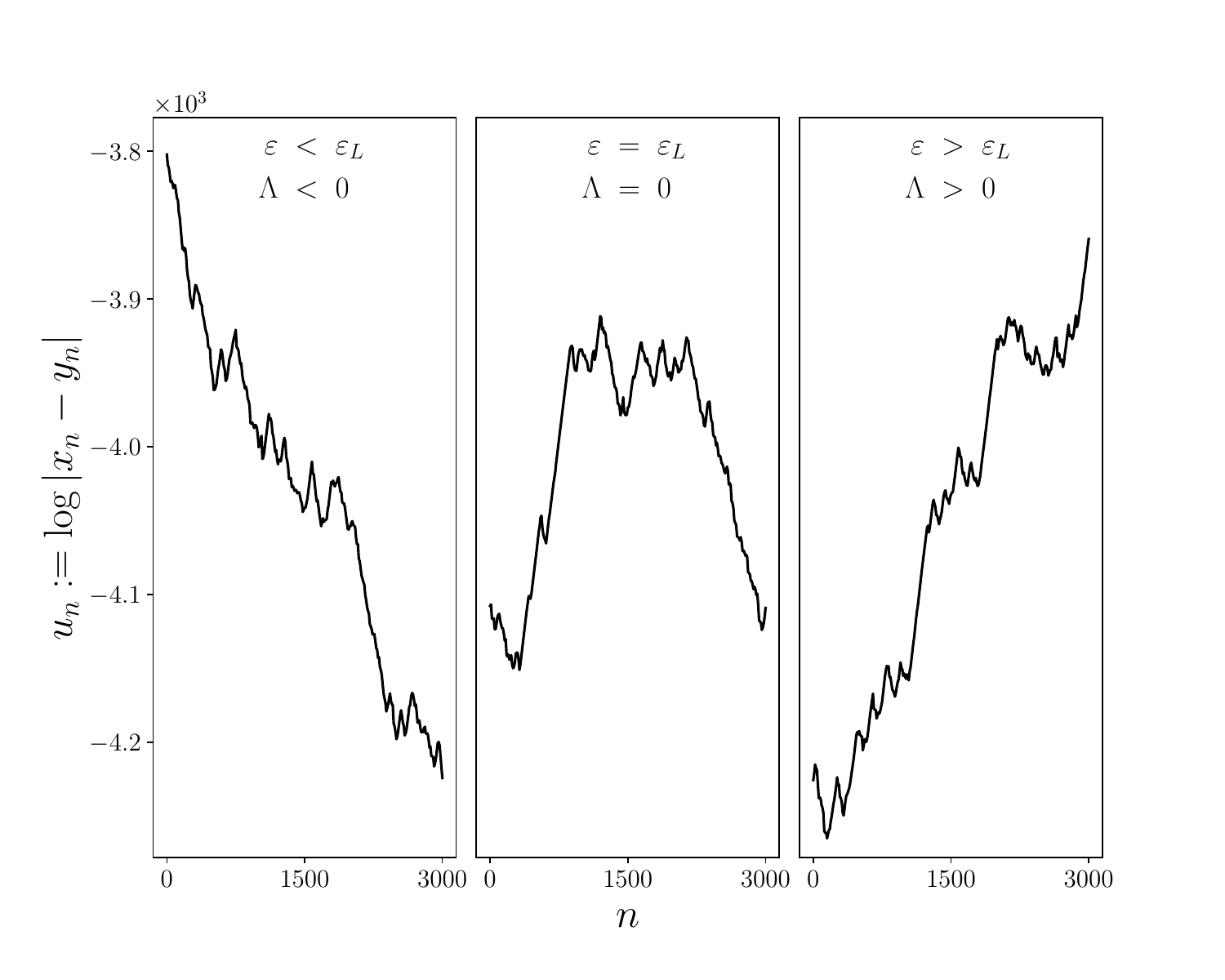}
    \caption{\label{fig:time-series}Time series evolution of $u_n$ for an initial condition $(x_0,y_0)$ of the two-point motion for phase II ($\e_D < \e = 0.00125 < \e_L, \Lambda <0$), at the transition point ($\e = 0.00146 \simeq \e_L, \Lambda = 0$), and phase III ($\e_L < \e = 0.00175, \Lambda >0$).}
\end{figure}

In FIG.~\ref{fig:time-series} we illustrate the typical evolution of $u_n$ for a pair of close initial conditions and for three representative values of $\e$: one in phase II ($\e_D<\e<\e_L,  \Lambda <0)$, one at the transition point $(\e = \e_L, \Lambda = 0)$, and one in phase III ($\e > \e_L, \Lambda > 0)$. On each time series, characteristic bursts stand out as finite-time trajectories where $u_n$ grows or decays linearly for relatively long time intervals, representing desynchronising and synchronizing excursions, respectively. Moreover, we observe in each time-series that desynchronising bursts have similar growth rates and, likewise, desynchronising ones have similar decay rates.

\section{A conditioned random dynamics perspective}

We view the noise-induced transition to chaos as characterized by the change of sign of the Lyapunov exponent due to a competition between stable and unstable regimes. This leads us to propose a compartmental model, partitioning the attractor $A$ into two regions, $M_{\Exp}$ and $M_{\Cont}$, with predominantly expanding and contracting dynamics, respectively. 
Each compartment is characterized by its own expected escape time, $\tau_\Exp$ and $\tau_\Cont$, as well as its effective exponential rate of expansion and contraction, $\Lambda_\Exp$ and $\Lambda_\Cont$. Thus, the overall exponential growth rate of the model can be expressed as the weighted average
\begin{equation}\label{eq:Lyap_model}
    \overline{\Lambda} \coloneqq \dfrac{{\tau}_{\Exp} \Lambda_{\Exp} + {\tau}_{\Cont}  \Lambda_{{\Cont}}}{{\tau}_{\Exp} + {\tau}_{\Cont}}.
\end{equation}
If the two-state model correctly reflects the dynamics of the random logistic map in Eq.~\eqref{eq:rlm}, then of course $\overline{\Lambda}\simeq \Lambda$. 

We note that a compartmental point of view towards noise-induced chaos, and Eq.~\eqref{eq:Lyap_model}, were also proposed in Ref.~\onlinecite{Lai2003}, cf.~also Ref.~\onlinecite{Lai2011}. The crucial difference between their treatment and ours is that the former relies on properties of transient dynamics in the deterministic limit. In particular, growth rates are identified with local Lyapunov exponents of attractors and non-attracting chaotic sets, which are established in a heuristic way, supported by numerical observations. Expected escape times are related to phenomenological quasipotentials, in a perturbative - small noise - setting.
In contrast, in this paper we identify the escape times and exponential growth rates in Eq.~\eqref{eq:Lyap_model} with expected escape times\cite{Ferrari1992} and conditioned Lyapunov exponents\cite{Engel2019, Castro2022} from the theory of conditioned random dynamical systems.\cite{Breyer1999, Collet2013} These are mathematically precise and do not rely on small noise assumptions, nor refer to deterministic limits.

The theory of conditioned random dynamics concerns the behavior of trajectories that stay within a domain $M$ of the state space for asymptotically long times. The corresponding statistics are described by the so-called quasi-stationary density $m$, which is the leading eigenmode of the conditioned transfer operator $\mathcal L_M$:\cite{Collet2013}
\begin{equation}\label{eq:transfer}
\mathcal L_M m \coloneqq \mathbb E (m \circ f_\w \cdot \mathbf{1}_M\circ f_\w) = \lambda_M m,
\end{equation}
where the expectation $\mathbb E$ is taken with respect to the noise realizations $\w$, $g\circ f_\w$ is the evolution of the density under the dynamics, $\mathbf{1}_M \circ f_\w$ controls survival of trajectories within $M$, and $\lambda_M \in (0,1)$ is the largest eigenvalue of $\mathcal L_M$. 
Indeed, the conditioned time evolution of any initial density converges - up to scalar multiple - to the quasi-stationary density $m$.\cite{Collet2013}

The eigenvalue $\lambda_M$ is the so-called (asymptotic) escape rate from $M$ since it follows from Eq.~\eqref{eq:transfer} that
\begin{equation}\label{eq:escape}
\lambda_M=\int_M \P(f_\w(x)\in M)m(x)\d x,
\end{equation}
i.e. $\lambda_M$ expresses the probability that trajectories remain in $M$ given that their initial conditions are distributed according to $m$. The associated expected escape time from $M$ is given by\cite{Ferrari1992}
\begin{equation}\label{eq:times}
     {\tau}_M = \sum_{n= 1}^{\infty} n (1-\lambda_M)\lambda_M^{n-1} = \frac{1}{1-\lambda_M},
\end{equation}
where $(1-\lambda_M)\lambda_M^{n-1}$ is the average over $m$ of the probability that a trajectory leaves $M$ for the first time at the $n$-th iterate.

The theory of conditioned random dynamics admits a version of Birkhoff's ergodic theorem involving averages over so-called quasi-ergodic, rather than quasi-stationary, densities.\cite{Darroch1965, Breyer1999, Castro2021} A probability density $q$ is quasi-ergodic if in the limit $n \to \infty$, the expectation of the time average for any observable $g$ conditioned on survival on $M$ equals the expectation of $g$ with respect to $q$. In particular, the conditioned Lyapunov exponent $\Lambda_M$ for the random logistic map in Eq.~\eqref{eq:rlm} is given by
\[%\label{eq:cond_lyap}
    \Lambda_M \coloneqq \int_M\ln|a(1-2y)|q(y) \d y.
\]
This conditioned Lyapunov exponent represents the infinitesimal exponential growth rate of orbits that remain asymptotically long within $M$. Like Lyapunov exponents, conditioned Lyapunov exponents provide useful dynamical insights.\cite{Engel2019}

The formal relationship between the quasi-stationary and quasi-ergodic densities is given by (see e.g. Ref.~\onlinecite[Theorem C]{Castro2021})
\begin{equation}\label{eq:qed}
    q(x) = \dfrac{v(x) m(x)}{\int_M v(y)m(y)\d y},
\end{equation}
where $v$ is the unique non-negative continuous eigenfunction of the adjoint operator of $\mathcal L_M$ with eigenvalue $\lambda_M$:
\[%\label{eq:conformal}
    \lambda_M v(x) = \int_M v(y)\mathcal \P(f_{\w}(x) \in \d y).
\]

In practice, quasi-stationary and quasi-ergodic distributions rarely admit analytic expressions, but numerical approximations of these densities can be obtained using Ulam's method.\cite{Ulam1960, Li1976, Bose2014} This entails a coarse-grained approximation of the conditioned transfer operator on a $k$-component partition of $M$, represented by a $k\times k$ sub-stochastic transition matrix $\mathbf{P}$, where the matrix elements $\mathbf{P}_{ij}$ denote the probability to evolve from the $i$-th to the $j$-th component. The maximal eigenvalue ${\lambda}$ of $\mathbf{P}$ approximates the escape rate $\lambda_M$ in Eq.~\eqref{eq:escape} as $k\to \infty$. The left probability eigenvector $\mathbf{m} = (m_1, \dots, m_k)$ of $\mathbf{P}$ associated with ${\lambda}$ approximates the quasi-stationary density $m$, and the right probability eigenvector $\mathbf{v} = (v_1, \dots, v_k)$ of $\mathbf{P}$ associated with ${\lambda}$ approximates the function $v$.\cite{Darroch1965} Therefore, the quasi-ergodic density $q$ of the conditioned process in Eq.~\eqref{eq:qed} is approximated by the probability vector $\mathbf{q} = (q_1,\dots, q_k)$ with
\[%\label{eq:qed_vector}
     q_i  = \dfrac{v_i m_i}{\sum_{j=1}^k v_j m_j}, \quad i = 1,\dots, k.
 \]

Returning to our two-state model, with expanding and contracting compartments $M_\Exp$ and $M_\Cont$, we propose to approximate the escape times $\tau_\Exp$ and $\tau_\Cont$ with the expected asymptotic escape times $\tau_{M_\Exp}$ and $\tau_{M_\Cont}$ obtained from the respective quasi-stationary densities, cf.~Eq.~\eqref{eq:times}. Similarly, we approximate the rates of expansion and contraction $\Lambda_\Exp$ and $\Lambda_\Cont$ with the respective conditioned Lyapunov exponents. In other words,
\[%\label{eq:chars}
\begin{array}{cccccc}
\tau_\Exp &\coloneqq& \tau_{M_\Exp}, & \quad \tau_\Cont &\coloneqq& \tau_{M_\Cont},\\
\Lambda_\Exp &\coloneqq& \Lambda_{M_\Exp}, & \quad \Lambda_\Cont &\coloneqq& \Lambda_{M_\Cont}.
\end{array}
\]
This choice implies the assumption that conditioned long-term asymptotic behavior dominates the dynamics in each compartment, neglecting short-term transients to this asymptotic behavior.

Inspired by the three-periodic attractor of the underlying deterministic logistic map, we consider the partition of the attractor $A$ into the following compartments:
\[%\label{eq:M_cont_exp}
\begin{array}{lcl}
    M_\Exp &\coloneqq& \{x \in A \,|\, |(f^3_\w)^{\prime}(x)| \geq 1,\, \forall \,\w\},\\
    M_{\Cont} &\coloneqq& \{x \in A \,|\, \exists\, \w \text{ s.t. } |(f^3_\w)^{\prime}(x)| < 1\}.
\end{array}
\]
For random dynamical systems like the logistic map in Eq.~\eqref{eq:rlm}, and with this type of compartments (that do not contain the support of any stationary density), it is known\cite{Castro2021} that unique quasi-stationary and quasi-ergodic densities exist, so all proposed expected escape times and conditioned Lyapunov exponents are well-defined.

Importantly, the theory of conditioned random dynamics applies equally to the bounded and unbounded noise settings. In particular, this theory can in principle also be used to give the numerical observations on transient dynamics for the random logistic map with unbounded Gaussian noise in Ref.~\onlinecite{Crutchfield1982} a rigorous mathematical footing.

\section{Noise-induced chaos}

\begin{figure}[htb]
    \centering
    \includegraphics[width = .48\textwidth]{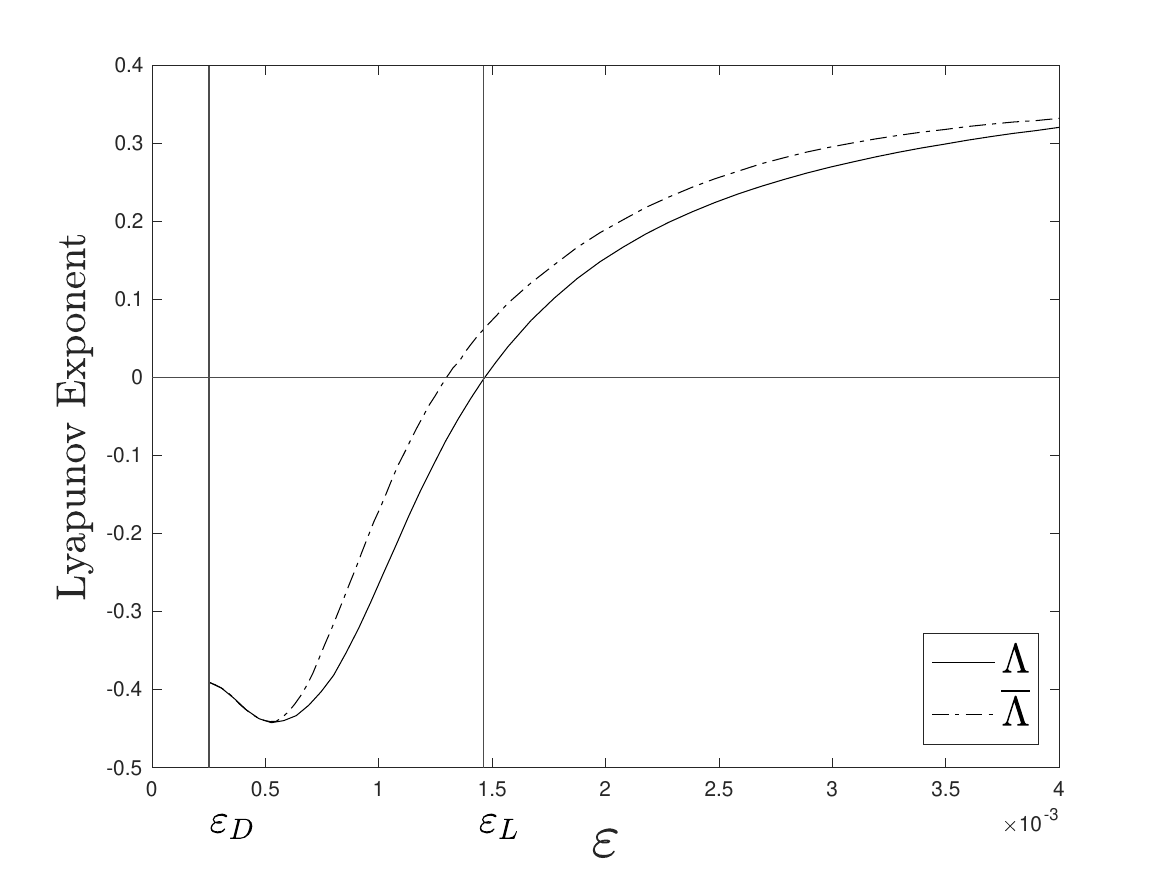}
    \caption{\label{fig:Lyaps_all} Lyapunov exponent $\Lambda$ of Eq.~\eqref{eq:rlm} as a function of noise amplitude obtained from the stationary density, and $\overline{\Lambda}$ obtained from the quasi-stationary and quasi-ergodic densities.}
\end{figure}

The numerically obtained Lyapunov exponent $\overline{\Lambda}$ of the corresponding two-state model is presented in FIG.~\ref{fig:Lyaps_all} for a range of noise amplitudes $\e$ increasing from phase II to phase III. For comparison, the Lyapunov exponent $\Lambda$ of the random logistic map is also depicted. We obtain excellent qualitative and good quantitative agreement. We discuss the reason for the slight overestimation of $\Lambda$ by $\overline{\Lambda}$ later in this section, but first focus on the insights we obtain about the noise-induced transition to chaos from the compartmental model.

\begin{figure}[bth]
    \centering    
    \includegraphics[width = .5\textwidth, trim=5.5cm 0cm 2cm 0cm, clip]{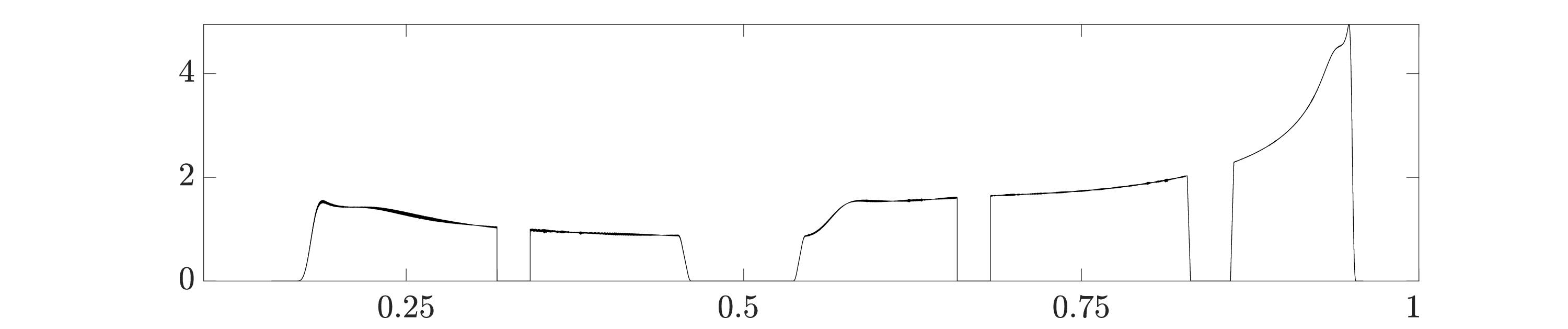}
    \footnotesize{(a) quasi-stationary density, $m$}
    \includegraphics[width = .5\textwidth, trim=5.5cm 0cm 2cm 0cm, clip]{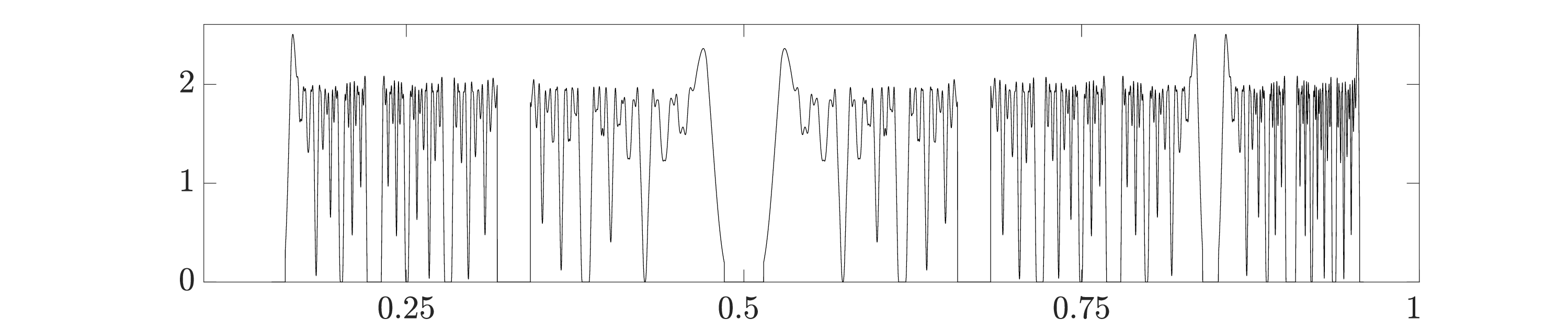}
     \footnotesize{(b) right eigenfunction, $v$}
    \includegraphics[width = .5\textwidth, trim=5.5cm 0cm 2cm 0cm, clip]{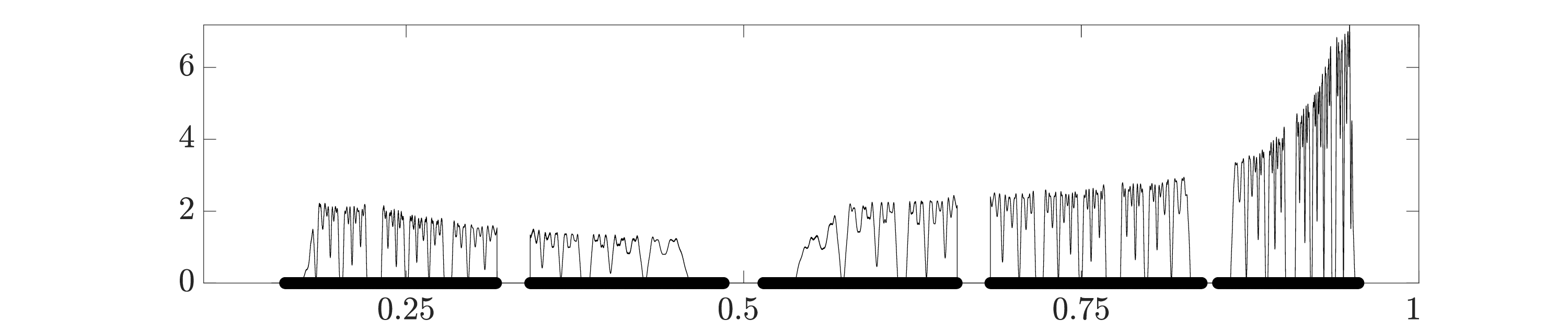}
     \footnotesize{(c) quasi-ergodic density, $q$}
    \caption{\label{fig:q_measures} From top to bottom, the quasi-stationary density $m$, the function $v$ and the quasi-ergodic density $q$, on the expanding region $M_{\Exp}$ for $\e = 0.00125$. These have been obtained using Ulam's method with $2^{14}$ components. The compartment $M_{\Exp}$ has been highlighted in bold on the bottom horizontal axis.}
\end{figure}

$\overline{\Lambda}$ is calculated from the expected escape times and conditioned Lyapunov exponents in the compartments $M_\Exp$ and $M_\Cont$, which in turn are obtained through the quasi-stationary and quasi-ergodic densities. For illustration and comparison, these densities are depicted in FIG.~\ref{fig:q_measures} for $M_\Exp$ and noise amplitude $\e = 0.00125$ (in phase II). The function $v$ that provides the connection between these densities, cf.~Eq.~\eqref{eq:qed}, is also given. 

\begin{figure}[hbt]
    \centering
    \includegraphics[width = .48\textwidth]{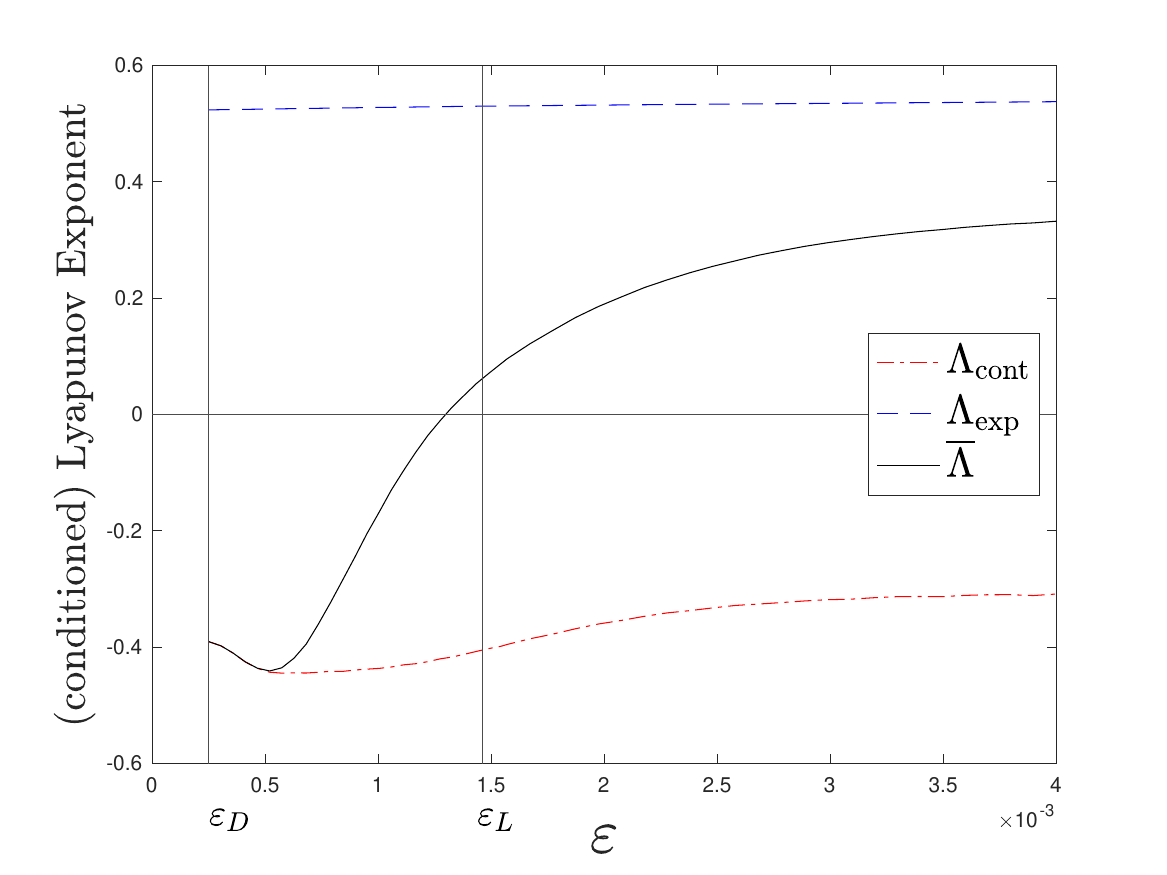}
    \caption{Conditioned Lyapunov exponent on the expanding and contracting components, $\Lambda_\Exp$ and $\Lambda_\Cont$, as a function of $\e$. The Lyapunov exponent $\overline{\Lambda}$ is presented to illustrate the role of the escape times in Eq.~\eqref{eq:Lyap_model} as $\Lambda_\Exp$ and $\Lambda_\Cont$ remain almost constant.}
    \label{fig:CLE}
\end{figure}

\begin{figure}[htb]
    \centering
    \includegraphics[width = .48\textwidth]{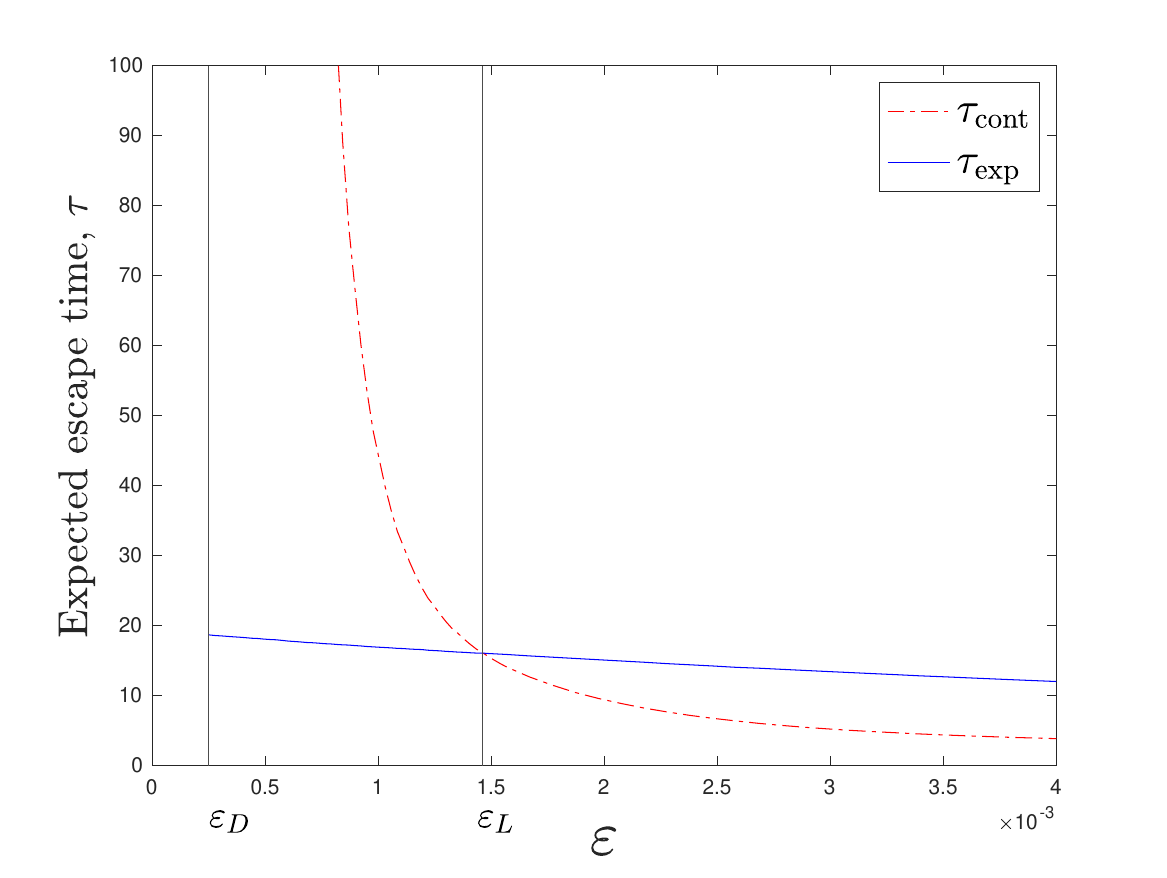}
    \caption{Expected escape times ${\tau}_\Exp$ and ${\tau}_{\Cont}$ from the expanding and contracting regions, respectively, as a function of the noise strength $\e$. Note that the intersection of $\tau_\Exp$ and $\tau_\Cont$ near the transition point $\e_L$ is not an indicator of phase change.}
    \label{fig:escape-times}
\end{figure}

The behavior of the Lyapunov exponent as a function of $\varepsilon$ can be understood from the corresponding behavior of the constituents of $\overline{\Lambda}$ in Eq.~\eqref{eq:Lyap_model}. As the noise strength $\varepsilon$ increases from phase II to phase III we track the conditioned Lyapunov exponents $\Lambda_\Exp$ and $\Lambda_\Cont$ in FIG.~\ref{fig:CLE}, as well as the expected escape times ${\tau}_\Exp$ and ${\tau}_\Cont$ in FIG.~\ref{fig:escape-times}.
We observe in FIG.~\ref{fig:CLE} that the growth and decay rates of the compartmental model are almost unchanged as a function of the noise amplitude, compared to the Lyapunov exponent $\overline{\Lambda}$. In FIG~\ref{fig:escape-times} we observe minor variability of ${\tau}_\Exp$ and fast decay of $\tau_{\Cont}$. Thus, with $\tau_\Exp$, $\Lambda_\Exp$ and $\Lambda_\Cont$ effectively constant, it is the significant decay of the expected escape time from the contracting region $\tau_\Cont$ that drives the noise-induced transition to chaos.

We conclude this section with a discussion of the slight overestimation of the Lyapunov exponent $\Lambda$ by $\overline{\Lambda}$. This is ultimately a consequence of the fact that the dynamics in each component are not entirely dominated by asymptotic conditioned behavior.
It turns out that the expected escape time from $M_\Cont$ and effective contraction rate in this compartment are well approximated by $\tau_\Cont$ and $\Lambda_\Cont$, respectively. The accuracy of the corresponding approximations by $\tau_\Exp$ and $\Lambda_\Exp$ in $M_\Exp$ are somewhat less reliable when the noise amplitude is not large. First of all, $\tau_\Exp$ is theoretically correct when the empirical distribution in $M_\Exp$ equals the quasi-stationary distribution. However, the true distribution is influenced by the way trajectories enter this region from $M_\Cont$, leading to an overestimation of escape times by $\tau_\Exp$. The appropriateness of the conditioned Lyapunov exponent $\Lambda_\Exp$ similarly relies on the closeness of the empirical distribution to the quasi-stationary distribution. In addition, it requires a sufficiently long escape time so that finite-time conditioned Lyapunov exponents at the time-scale of the expected escape time are close to the asymptotic conditioned Lyapunov exponent. We have observed in numerical experiments that, at the time scale of the expected escape time, the finite-time Lyapunov distribution is essentially bi-modal rather than entirely concentrated around $\Lambda_\Exp$. As the right-most peak of the bi-modal distribution lies around $\Lambda_\Exp$, $\Lambda_\Exp$ overestimates the real effective growth rate. All in all, inaccuracies in both $\tau_\Exp$ and $\Lambda_\Exp$ lead to a slight overestimation of $\overline{\Lambda}$, as observed in FIG.~\ref{fig:Lyaps_all}. 
The quantitative aspects of the compartmental model could be improved by taking account more accurately of the inhomogeneity of escape times and the actual distribution of finite-time conditioned Lyapunov exponents, rather than relying on asymptotic values. We envisage this to be achievable considering higher order corrections beyond the leading asymptotics.

\section{Conclusion and outlook}

In this paper, we have shown how the noise-induced transition to chaos in a random logistic map can be understood from a conditioned random dynamics point of view. In a two-compartment model, representing the competition between expanding and contracting behavior, we have found that the bifurcation is characterized by a fast decay of the escape time from the contracting compartment. We conjecture this feature to be universal in noise-induced transitions of this type, and have indeed observed this also in other examples, such as the two-dimensional Hénon map with bounded additive noise in the period 7 window.\cite{KongMing2014} A full treatment of such higher dimensional examples, analogous to the random logistic map, is in principle possible, but requires substantially more numerical effort. Firstly, bifurcations from phase I to phase II in such examples require more analysis from the set-valued point of view.\cite{Tey2022} Secondly, the computation of conditioned Lyapunov exponents relies on estimating the quasi-ergodic density in the tangent bundle of compartments.\cite{Castro2022} Examples of noise-induced chaos in SDEs (with unbounded noise) involve similar challenges for the determination of conditioned Lyapunov exponents, as chaotic dynamics in such systems can arise only in dimension two or higher.\cite{Crauel1998} We intend to report on higher dimensional examples in forthcoming publications.

Long transients near deterministic repellers have been previously proposed as an essential ingredient for noise-induced chaos in the small noise regime.\cite{Tel2008, Lai2011} However, from our case study it transpires that small noise is not essential, nor sufficient, to explain noise-induced chaos. 
Indeed, while ``stickiness'' of the repelling region may well be related to its deterministic limit, crucially, noise needs to be sufficiently large in order to facilitate expedient escape from the attracting region, enabling the transition from negative to positive Lyapunov exponent. Moreover, we have observed noise-induced chaos to arise also in the random logistic map Eq.~\eqref{eq:rlm} with noise amplitude $\e=0.074$ and parameter value $a=3.4$ (between the first and second period-doubling bifurcation in the deterministic logistic map, where there are no chaotic transients).

Conditioned random dynamics is also useful to elucidate the understanding of chaotic transients in deterministic systems by considering the zero noise limit.\cite{Bassols2023} Indeed, the quasi-ergodic measures that underlie conditioned Lyapunov exponents are reminiscent of the ``natural measures'' on repellers introduced in Ref.~\onlinecite{Kantz1985} (see Fig.~4 therein), which have been effectively used to define and compute local Lyapunov exponents\cite{Lai2003, Lai2011}. While quasi-ergodic measures in random systems have a solid mathematical foundation\cite{Darroch1965, Breyer1999, Castro2021}, in the context of long chaotic transients, natural measures on repellers are defined with reference to numerical explorations.\cite{Kantz1985, Lai2011} It turns out\cite{Bassols2023} that quasi-ergodic measures near random hyperbolic repellers converge to such natural invariant measures in the zero noise limit, and closely relate to invariant measures for deterministic dynamical systems with holes.\cite{Liverani2003}

\begin{acknowledgements}
    The authors gratefully acknowledge support from the EPSRC Centre for Doctoral Training in Mathematics of Random Systems: Analysis, Modelling and Simulation (EP/S023925/1). JSWL thanks IRCN (Tokyo) and GUST (Kuwait) for their research support. We would like to thank M.~M.~Castro, H.~Chu, E.~Gibson, V.~P.~H.~Goverse, T.~Pereira, M.~Rasmussen, Y.~Sato, M.~Tabaro, G.~Tenaglia and D.~Turaev for useful discussions.
\end{acknowledgements}

\section*{AUTHOR DECLARATION}
\subsection*{Conflict of Interest} The authors have no conflicts to disclose.
\subsection*{Data Availability}
The data and code that support the findings of this paper are openly available in \url{https://github.com/Bernat-BC/Ni-Chaos}.
\subsection*{Author Contributions}
\textbf{Bernat Bassols-Cornudella:} Conceptualization (equal); software and visualizations (lead); writing (equal). \textbf{Jeroen S.W. Lamb:} Conceptualization (equal); writing (equal).
\section*{References}
\bibliography{references}
\end{document}